\documentclass{PoS}

\def\beq{\begin{equation}}
\def\eeq{\end{equation}}
\def\beqa{\begin{eqnarray}}
\def\eeqa{\end{eqnarray}}

\title{Two-loop resummation for QCD hard scattering}

\ShortTitle{Two-loop resummation for QCD hard scattering}
 
\author{\speaker{Nikolaos Kidonakis}%
        \thanks{This work was supported by the National Science 
Foundation under Grant No. PHY 0855421.}\\
       Kennesaw State University, USA\\
       E-mail: \email{nkidonak@kennesaw.edu}}

\abstract{
I present results for the resummation of soft and collinear
gluon contributions to QCD hard-scattering cross sections at two loops.
This requires the calculation of UV and IR divergences through two loops
and the construction of soft anomalous dimensions that control resummation at
NNLL. Explicit results are provided for processes that involve massless and/or
massive quarks, including single top and top pair production.
}

\FullConference{XVIII International Workshop on Deep-Inelastic Scattering and Related Subjects\\
                 April 19 -23, 2010\\
                 Convitto della Calza, Firenze, Italy}

\begin{document}

\section{Resummation}

Soft-gluon corrections are important in many QCD processes, particularly near threshold.
The calculation of these corrections is needed at higher orders for increased accuracy in theoretical predictions.
Terms of the form  $[\ln^k(s_4/M^2)/s_4]_+$, where $s_4\rightarrow 0$ at 
threshold, arise from incomplete cancellations of infrared divergences between virtual diagrams and real diagrams 
with soft (low-energy) gluons.
These soft-gluon corrections exponentiate.  The resummation follows from factorization properties of the cross section. 
At next-to-leading-logarithm (NLL) accuracy resummation requires one-loop calculations in the eikonal approximation; 
at next-to-next-to-leading-logarithm (NNLL) accuracy it requires two-loop calculations.
 
The resummation is performed in moment space, with $N$ the moment variable conjugate to $s_4$. 
We use renormalization group evolution (RGE) to evolve the soft-gluon function associated with soft-gluon emission. 
The resummed cross section is 
\beq
{\hat{\sigma}}^{res}(N)=
\exp\left[E(N)\right] \; {\rm tr} \left\{H\left(\sqrt{s}\right)
\exp \left[\int_{\sqrt{s}}^{{\sqrt{s}}/{\tilde N}}
\frac{d\mu}{\mu} \; \Gamma_S^{\dagger}(\mu)\right] \;
S\left(\frac{\sqrt{s}}{\tilde N}\right) \;
\exp \left[\int_{\sqrt{s}}^{{\sqrt{s}}/{\tilde N}}
\frac{d\mu}{\mu}\; \Gamma_S (\mu)\right]\right\} 
\label{resHS}
\eeq
where the exponent $E$ resums universal collinear and soft gluon emission 
\cite{GS87,CT89} from 
the incoming and outgoing partons, 
$H$ is the $N$-independent hard-scattering function, and $S$ is the  soft-gluon 
function whose RGE is governed by $\Gamma_S$, the soft anomalous dimension 
\cite{NKGS,NKNNNLO}. The functions $H$, $S$, and $\Gamma_S$ are matrices in color-exchange space.

We calculate $\Gamma_S$ in the eikonal approximation, in which the 
Feynman rules for the quark-gluon vertex give 
$g_s T_F^c \, (v^{\mu}/v\cdot k)$, with $g_s^2=4\pi\alpha_s$, 
$v$ a dimensionless velocity vector, $k$ the gluon momentum, and $T_F^c$ the generators of SU(3).
We perform the calculations in momentum space and Feynman gauge. 
We expand the soft anomalous dimension as $\Gamma_S=(\alpha_s/\pi) \Gamma_S^{(1)}+(\alpha_s/\pi)^2 \Gamma_S^{(2)}+\cdots$, 
and determine it through two loops for a number of processes.

\section{Massive soft (cusp) anomalous dimension} 

\begin{figure}
\begin{center}
\includegraphics[width=0.5\textwidth]{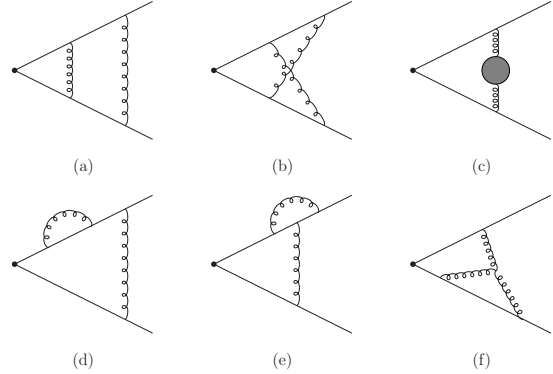}
\caption{Two-loop vertex-correction graphs.}
\label{loopg2}
\end{center}
\end{figure}

We begin with  two-loop results for the soft (cusp \cite{KorRad}) anomalous dimension for 
$e^+ e^- \rightarrow t {\bar t}$ \cite{NK2l}.
Calculations with massive quarks present considerable complications in addition to 
those with massless quarks \cite{ADS}.
The two-loop vertex-correction eikonal diagrams are shown in Fig. 1; there are additional diagrams with top-quark 
self energies \cite{NK2l}.
We include counterterms for all graphs and multiply with corresponding color factors.
Ultraviolet (UV) and infrared (IR) singularities are separated, and 
we determine the two-loop soft anomalous dimension from the UV poles of the sum of the 
graphs. In terms of the cusp angle \cite{KorRad} 
$\gamma=\ln[(1+\beta)/(1-\beta)]$, with $\beta=\sqrt{1-4m^2/s}$ and $m$ the heavy quark mass, 
we get $\Gamma_S^{(1)}=C_F (\gamma \coth\gamma-1)$  and \cite{NK2l}
\beqa
\Gamma_S^{(2)}&=&\frac{K}{2} \, \Gamma_S^{(1)}
+C_F C_A \left\{\frac{1}{2}+\frac{\zeta_2}{2}+\frac{\gamma^2}{2}
-\frac{1}{2}\coth^2\gamma\left[\zeta_3-\zeta_2\gamma-\frac{\gamma^3}{3}
-\gamma \, {\rm Li}_2\left(e^{-2\gamma}\right)
-{\rm Li}_3\left(e^{-2\gamma}\right)\right] \right.
\nonumber \\ && \hspace{25mm} \left.
{}-\frac{1}{2} \coth\gamma\left[\zeta_2+\zeta_2\gamma+\gamma^2
+\frac{\gamma^3}{3}+2\, \gamma \, \ln\left(1-e^{-2\gamma}\right)
-{\rm Li}_2\left(e^{-2\gamma}\right)\right] \right\}
\label{Gamma2}
\eeqa
where 
$K=C_A (67/18-\zeta_2)-5n_f/9$.
$\Gamma_S^{(2)}$ vanishes at 
$\beta=0$, the threshold limit, and diverges at $\beta=1$, 
the massless limit. 
In the case where one quark is massive and one is massless the result simplifies and we find 
$\Gamma_S^{(2)}=(K/2) \Gamma_S^{(1)}+C_F C_A (1-\zeta_3)/4$.

\section{Single top quark processes}

The color structure gets more complicated with more than two colored partons 
in a hard-scattering process. The cusp anomalous dimension, presented in the previous section, 
is an essential component of calculations for other processes.
Next, we compute two-loop soft anomalous dimensions for single top production via the $s$-channel,  
and for associated production of a top quark with a $W$ boson or a charged Higgs.

\subsection{$s$-channel single top quark production}

For $s$-channel single-top production the diagrams involve different pairings of the 
two-loop diagrams for the cusp anomalous dimension. We find the one-loop result for the soft anomalous dimension 
\cite{NKst,NKs}   
\beq
\Gamma_{S,\, {\rm top \; s-ch}}^{(1)}=C_F \left[\ln\left(\frac{s-m_t^2}{m_t\sqrt{s}}\right)
-\frac{1}{2}\right] 
\eeq
and the two-loop \cite{NKs} expression 
\beq
\Gamma_{S,\, {\rm top \; s-ch}}^{(2)}=\frac{K}{2} \Gamma_{S,\, {\rm top \; s-ch}}^{(1)}
+C_F C_A \frac{(1-\zeta_3)}{4} \, .
\eeq

Using this result in Eq. (\ref{resHS}) and expanding the resummed cross section to next-to-next-to-leading 
order (NNLO) in the strong coupling, we derive an approximate NNLO cross section. 
As shown in \cite{NKs,POSDIS} the NNLO approximate corrections enhance the NLO cross section by about 15\%  
at the Tevatron and about 13\% at LHC energies.

\subsection{Associated production of a top quark with a $W^-$ or $H^-$}

\begin{figure}
\begin{center}
\includegraphics[width=0.31\textwidth]{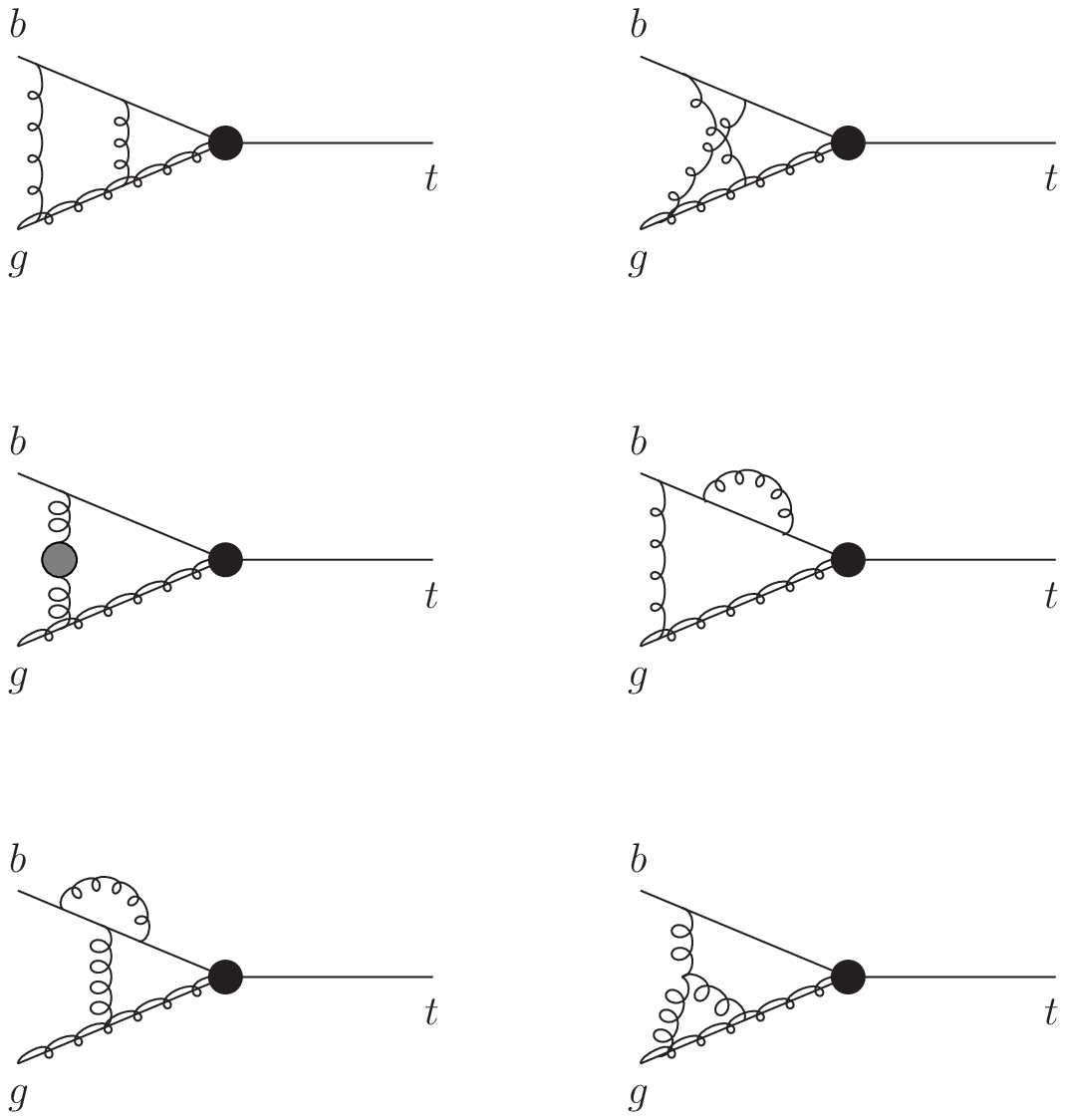}
\hspace{3mm}
\includegraphics[width=0.31\textwidth]{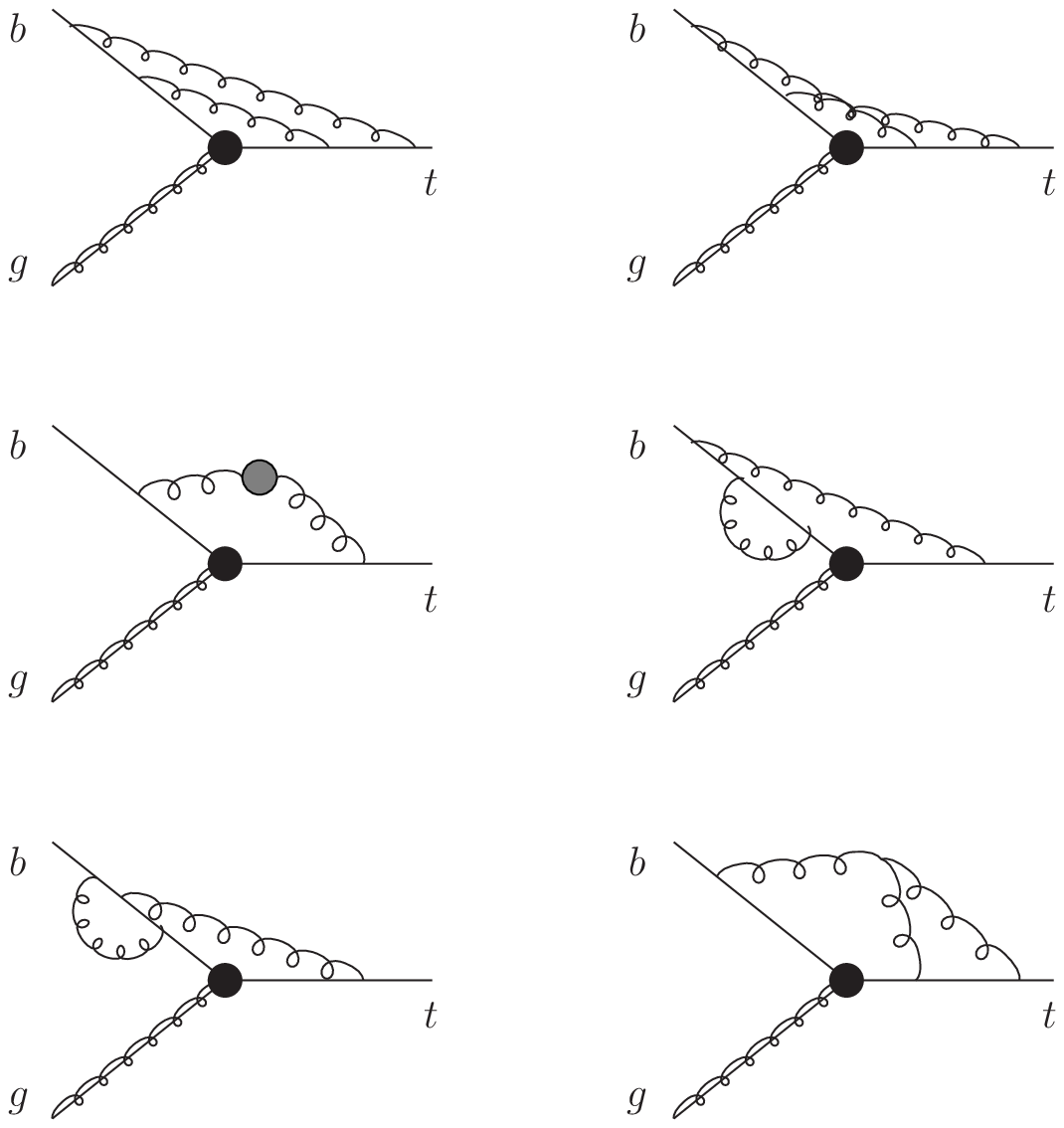}
\hspace{3mm}
\includegraphics[width=0.31\textwidth]{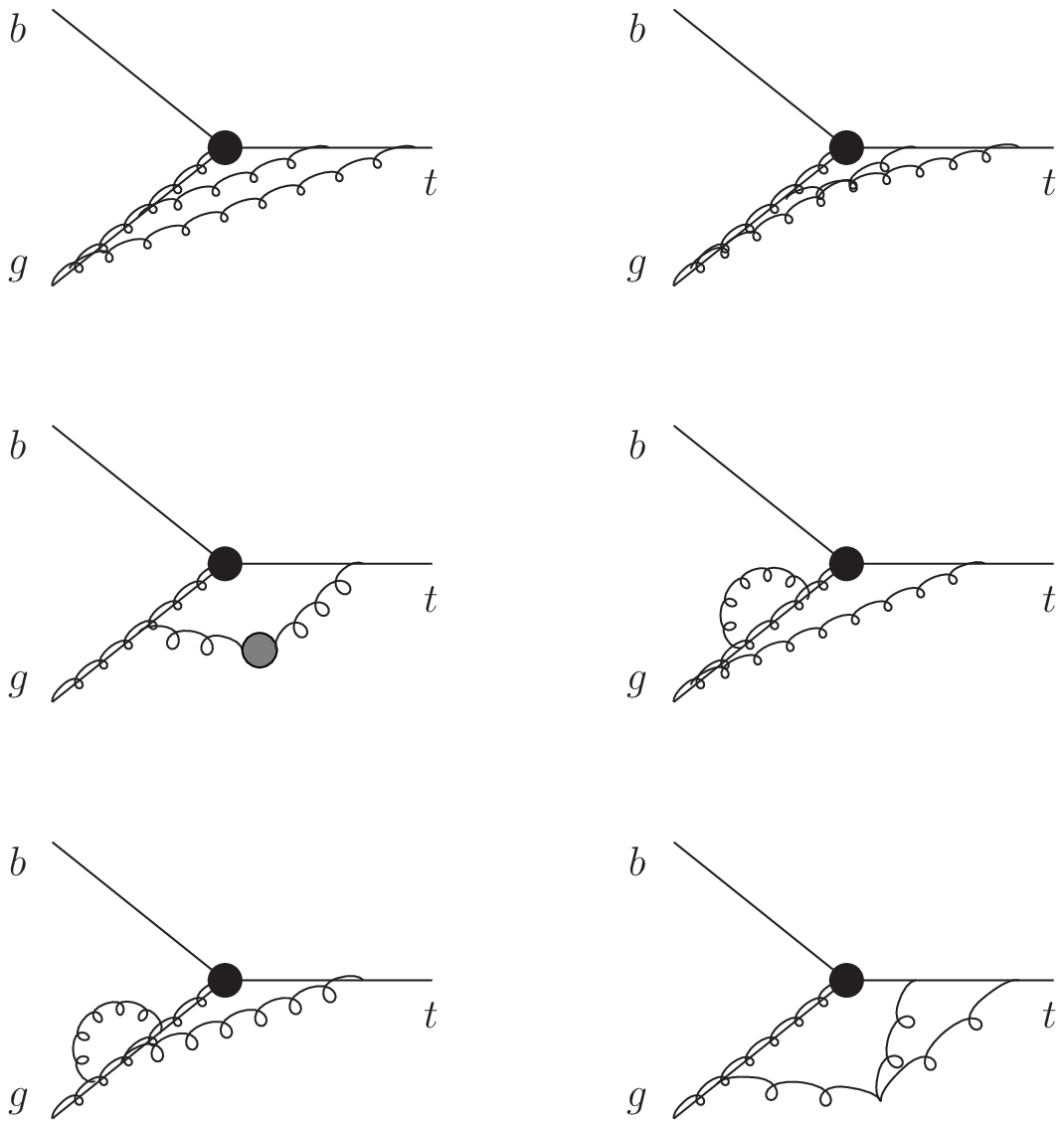}
\caption{Eikonal two-loop graphs for $tW^-$ or $tH^-$ production.}
\label{eiktW}
\end{center}
\end{figure}

For the processes $bg \rightarrow tW^-$ or $bg \rightarrow tH^-$ the two-loop 
vertex-correction eikonal diagrams are shown in Fig. 2 (there are additional 
top-quark self-energy graphs not shown here). Note that graphs with gluons attached to all three eikonal 
lines, not shown here, either vanish or do not contribute to the two-loop soft anomalous dimension. 
The one-loop result for the soft anomalous dimension for $bg \rightarrow tW^-$ is
\beq
\Gamma_{S,\, tW^-}^{(1)}=C_F \left[\ln\left(\frac{m_t^2-t}{m_t\sqrt{s}}\right)
-\frac{1}{2}\right] +\frac{C_A}{2} \ln\left(\frac{m_t^2-u}{m_t^2-t}\right)
\eeq
and at two loops we find 
\beq
\Gamma_{S,\, tW^-}^{(2)}=\frac{K}{2} \Gamma_{S,\, tW^-}^{(1)}
+C_F C_A \frac{(1-\zeta_3)}{4} \, .
\eeq
The NNLO approximate corrections for $tW^-$ production enhance the NLO cross section by about 8\%  
at LHC energies \cite{POSDIS}.

The same analytical result for $\Gamma_S$ holds for $bg\rightarrow tH^-$ \cite{POSDIS,NKchiggs}.
The NNLO approximate corrections for $tH^-$ production provide around 15\% to 20 \% enhancement over NLO   
at LHC energies \cite{POSDIS}.

Related results can also be derived for direct photon production.

\section{Top-antitop pair hadroproduction}

Resummation for $t{\bar t}$ production at hadron colliders at NLL accuracy was presented in \cite{NKGS} where the 
one-loop soft anomalous dimension matrix was first calculated. Detailed 
numerical studies including the NLL terms and sets of subleading terms \cite{NKRV} 
have shown that threshold resummation increases the cross section of 
$t{\bar t}$ production and significantly reduces the scale dependence at both Tevatron and LHC energies. 

The two-loop results of \cite{NK2l} are an essential ingredient in 
determining the complete two-loop matrix result for $t{\bar t}$ hadroproduction. If we rewrite Eq. (\ref{Gamma2}) 
as  $\Gamma_S^{(2)}=(K/2)\Gamma_S^{(1)}+C_F C_A M_{\beta}$ (i.e. $M_{\beta}$ 
denotes the terms inside the curly brackets in Eq. (\ref{Gamma2}), with the cusp angle 
$\gamma$ expressed explicitly in terms of $\beta$), then 
the diagonal elements of the two-loop soft anomalous dimension matrix, for both $q{\bar q}\rightarrow t{\bar t}$ and 
$gg \rightarrow t{\bar t}$ channels, can be written 
as a multiple of the one-loop result (in Feynman gauge) plus a multiple of 
$M_{\beta}$. For example, in a singlet-octet basis 
for the $q{\bar q}\rightarrow t{\bar t}$ channel we have 
\beqa
\Gamma_{q{\bar q} \,11}^{(2)}&=&\frac{K}{2} \Gamma_{q{\bar q} \,11}^{(1)}
+C_F C_A \, M_{\beta}\, ,
\nonumber \\
\Gamma_{q{\bar q} \,22}^{(2)}&=&
\frac{K}{2} \Gamma_{q{\bar q} \,22}^{(1)}
+C_A\left(C_F-\frac{C_A}{2}\right) \, M_{\beta} \, .
\label{Gamma2qqtt}
\eeqa
Similar results are found for the $gg$ channel. The off-diagonal elements 
obey a similar relation in terms of a function $N_{\beta}$ which involves a 
subset of the terms in $M_{\beta}$. Complete results will appear in a 
forthcoming paper. It is important to note that the inclusion of the complete 
two-loop expressions for the soft anomalous dimension matrix do not change 
the results of \cite{NKRV} by more than {\emph{O}}(per mille), thus showing the 
robustness of those earlier results and proving that they captured the dominant
subleading terms as was explained in detail in \cite{NKRV}.

\end{document}